\documentstyle[eqsecnum,aps,12pt]{revtex}
\newcommand{\la}[1]{\label{#1}}
\newcommand{\be}{\begin{equation}}
\newcommand{\ee}{\end{equation}}
\newcommand{\ba}{\begin{eqnarray}}
\newcommand{\ea}{\end{eqnarray}}
\newcommand{\bi}{\begin{itemize}}
\newcommand{\ei}{\end{itemize}}

\newcommand{\nr}[1]{(\ref{#1})}

\newcommand{\bfp}{{\bf p}}
\newcommand{\bfx}{{\bf x}}

\newcommand{\RR}{{\rm I\kern -.2em  R}}
\newcommand{\eq}{Eq.~}

\newcommand{\se}{Sec.~}

\def\lsi{\raise0.3ex\hbox{$<$\kern-0.75em\raise-1.1ex\hbox{$\sim$}}}
\def\gsi{\raise0.3ex\hbox{$>$\kern-0.75em\raise-1.1ex\hbox{$\sim$}}}

\begin{document}

\begin{titlepage}
\begin{flushright}
CERN-TH/2002-136\\
hep-ph/0206209\\
\end{flushright}
\begin{centering}
\vfill

\mbox{\bf QCD EFFECTIVE ACTIONS}

\mbox{\bf FROM THE SOLUTIONS OF THE TRANSPORT EQUATIONS}

\vspace{0.9cm}

Cristina Manuel\footnote{Electronic address: {\tt
cristina.manuel@cern.ch}}

{\it Theory Division, CERN \\
CH - 1211 Geneva 23, Switzerland}

\vspace{0.5cm}

Stanis\l aw Mr\' owczy\' nski\footnote{Electronic address: {\tt
mrow@fuw.edu.pl}}

{\it So\l tan Institute for Nuclear Studies \\
ul. Ho\.za 69, PL - 00-681 Warsaw, Poland \\
and Institute of Physics, \'Swi\c etokrzyska Academy \\
ul. \'Swi\c etokrzyska 15, PL - 25-406 Kielce, Poland}

\date{19-th October 2002}

\vspace*{0.8cm}

\end{centering}

\noindent
We solve the collisionless transport equations of a quark-gluon
plasma interacting through mean chromodynamic fields. The system
is assumed to be translation invariant in one or more space-time
directions. We present exact solutions that hold if the vector
gauge fields in the direction of the translation invariance
commute with their covariant derivatives. We also solve the
equations perturbatively when the commutation condition is
relaxed. Further, we derive the color current and the associated
effective action. For the static quasi-equilibrium system, our
results reproduce the full one-loop effective action of QCD  in the
presence of constant background fields, where the above mentioned
commutation condition is  satisfied.
\vfill
\noindent

\noindent
PACS numbers: 12.38.Mh, 05.20.Dd, 11.10.Wx

\vspace*{1cm}

\noindent
CERN-TH/2002-136\\
June 2002, revised October 2002

\vfill

\end{titlepage}


\section{Introduction}


When the temperature $T$ of the quark-gluon plasma is much greater
than the QCD scale parameter $\Lambda_{\rm QCD}$, the hard modes,
i.e. those with momenta of the order of $T$ or larger, are weakly
interacting and they can be described within perturbative QCD
\cite{Col75}. The dynamics of the soft sector, however, remains
non-perturbative even at arbitrarily large temperature
\cite{Gross:1980br}, as signalled by severe infrared divergences
\cite{Lin80}. Then, one has to refer to effective theories to get
an insight into the soft mode dynamics. Such theories, see e.g.
\cite{Braaten:1989mz,Bod98}, can be derived from QCD by
integrating out the hard modes, but constructing them is by far
not a simple task. Consequently, one often relies on more or less
heuristic approaches, usually exploiting a semi-classical or
classical field approximation because the occupation numbers
of the soft gluonic modes are large.

A very natural effective approach is provided by the kinetic
theory, where the hard modes are treated as (quasi-)particles while
the soft gluonic ones contribute to the chromodynamic mean field.
The transport theory has been formulated in two versions. The
first one treats the color degrees of freedom as a classical
continuous variable which, as position or momentum, evolves in
time. A starting point of the theory are the Wong equations
\cite{Won70}, which describe a classical particle that interacts with
the chromodynamic field due to the color charge. Then, one
immediately gets the Liouville and the transport equations
\cite{Hei83} of a many-body quark-gluon system. The physical
content of the theory is rather transparent and numerous results,
for example transport coefficients, can be easily obtained. Even
the simplest collisionless transport equations, where the
dissipation phenomena are neglected, provides a surprisingly rich
dynamics. The transport theory with the classical color became
really reliable when the theory was found \cite{Kel94} to
reproduce the QCD hard-thermal-loop dynamics
\cite{Braaten:1989mz,Tay90,Bra92}. It was further established
\cite{Lit99} that the theory supplemented by the collision terms,
obtained integrating out soft fluctuations around the mean
fields, agrees with the QCD effective approaches
\cite{Bod98,Arn99}. The relationship between the transport theory
with classical color and QCD can be found through the study of the
quantum path integral within a saddle-point approximation
\cite{Jal00}.

In the second version of the QCD transport theory \cite{Hei83,Win84}, the
color charges are represented, in  full accordance with QCD, by a matrix
structure of the distribution function. The Vlasov transport equation of
quarks was derived \cite{Elz86a} directly from QCD, by analyzing the motion
of quantum quarks in the classical chromodynamic field. The gluon
transport equation was found \cite{Elz86b,Elz87}, by splitting the gluon
field into the mean field and the contribution representing the particle
excitations. It was further observed \cite{Mro89} that the quark and gluon
transport equations are formally identical when the first one is written
in the fundamental representation and the second one in the adjoint
representation. Then, the quark and gluon distribution functions are
$N_c \times N_c$ and $(N_c^2 -1) \times (N_c^2 -1)$ matrices, respectively,
for the $SU(N_c)$ gauge group. The early development of the quark-gluon
kinetic theory was summarized at ref. \cite{Elz89}.

In quasi-equilibrium, the matrix transport theory was proved \cite{Bla93}
to be fully equivalent to the QCD hard-loop approach
\cite{Braaten:1989mz,Tay90,Bra92}. The kinetic approach, which can be
treated as a local representation of the non-local hard-loop action,
is particularly useful to study the collective excitations of the quark-gluon
plasma, see \cite{Bla02} for a review. More recently, the quasi-equilibrium
kinetic equations have been derived beyond the collisionless limit
\cite{Bla99} and the QCD effective theories \cite{Bod98} have again been
correctly reproduced.

A natural question that arises is where the agreement between the
kinetic theory, either with the classical color or in the matrix
form, and the finite temperature diagrammatic approach breaks
down. Surprisingly enough, the effective action of the static
fields provided by the kinetic equations agrees with that
obtained within perturbative QCD even at the $g^3$ order
\cite{Bod01}, where an operator responsible for $C$-odd processes
appears for systems with finite baryon density. However, in
a subsequent study \cite{Laine:2001my} the kinetic theory with
the classical color has been found to reproduce the $g^4$
contribution to the effective action only in the limit of
high-dimensional color representations. Thus, the limitations
of the classical approach have been explicitly determined.

The aim of this paper is to clarify whether the difficulties faced
by the classical color transport theory can be overcome when the
matrix formulation is used. We explore how far the limits of the
non-Abelian kinetic approach can be extended. For this purpose, we
look for the solutions of the transport equations for quarks and
gluons interacting with a chromodynamic mean field. The system is
assumed to be translation invariant in one or more space-time
directions. Thus, our considerations hold, in particular, for static
and for homogeneous systems. We first present exact solutions when
the vector gauge fields in the direction of the translation invariance
commute with their covariant derivatives. Then, we use perturbation
theory to solve the transport equation when the commutation condition
is relaxed. Once the solutions are known, we derive the corresponding
color current, and then the associated effective action. In the case
of thermodynamic equilibrium, our results agree with those obtained
by computing the one-loop effective action of QCD in the presence
of a constant background field,
which also corresponds to the effective potential of the dimensionally
reduced theory \cite{Har00}. However, for static and non-constant
background fields such that the commutation condition is not satisfied,
we find additional non-local  operators that correct the effective
potential of \cite{Har00}, in what  seems to be a discrepancy between
the two approaches.

The paper is organized as follows. In \se \ref{sec-transport} we  review
the transport theory approach we use in this paper. Exact solutions of the
transport equations for translation invariant systems are discussed in
\se \ref{sec-exact} while in \se \ref{pert-sol} we solve the equations
perturbatively. The effective actions for static and for homogeneous systems
close to equilibrium are derived in \se \ref{actions} and we conclude
our considerations in \se \ref{discussion}. The evaluation of some
momentum integrals is left for  Appendix \ref{integrals} and we collect in
Appendix \ref{traces}  some formulas of the traces of $SU(N_c)$
generators in the adjoint representation. Finally, in Appendix C we 
briefly discuss the one-loop effective action 
\cite{Gross:1980br,Weiss:1980rj,cka} which is compared to our results.


\section{Transport Equations}
\label{sec-transport}


In this section we briefly review the transport theory of quarks and
gluons \cite{Elz89,Mro89}. While we will restrict the discussion to
QCD, with $N_f$ massless quarks and antiquarks carrying color
in the fundamental representation and gluons in the adjoint, the results
could easily be generalized  to a different non-Abelian
theory with different field content.

The distribution function of (anti-)quarks
$Q( p,x)\;\bigr(\bar Q( p,x)\bigl)$ is a hermitian
$N_c\times N_c$ matrix in color space (for a $SU(N_c)$ color
group); $x$ denotes the space-time quark coordinate and $ p$
its momentum, which is not constrained by the mass-shell condition.
The spin of quarks and gluons is taken into account as an internal
degree of freedom. The distribution function transforms under
a local gauge transformation $M$ as
\begin{equation}\label{Q-transform}
Q( p,x) \rightarrow M(x)Q( p,x)M^{\dag }(x) \;.
\end{equation}
Here and most cases below, the color indices are suppressed. The
distribution function of hard gluons is a hermitian
$(N_c^2-1)\times (N_c^2-1)$ matrix, which transforms as
\begin{equation}\label{G-transform}
G( p,x) \rightarrow {\cal M}(x)G( p,x){\cal M}^{\dag }(x) \;,
\end{equation}
where
$$
{\cal M}_{ab}(x) = {\rm Tr}\bigr[\tau_a M(x) \tau_b M^{\dag }(x)] \ ,
$$
with $\tau_a ,\; a = 1,...,N_c^2-1$ being the $SU(N_c)$ group generators
in the fundamental representation with ${\rm Tr} (\tau_a \tau_b) = \frac12
\delta_{ab}$.

In a collisionless limit, the distribution functions of quarks and gluons
satisfy the transport equations:
\begin{mathletters}
\label{transport}
\begin{eqnarray}
p^{\mu} D_{\mu}Q( p,x) + {g \over 2}\: p^{\mu}
\left\{ F_{\mu \nu}(x),
{\partial Q( p,x) \over \partial p_{\nu}}\right\} &=&  0\;,
\label{transport-q}  \\
p^{\mu} D_{\mu}\bar Q( p,x) - {g \over 2} \: p^{\mu}
\left\{ F_{\mu \nu}(x),
{\partial \bar Q( p,x) \over \partial p_{\nu}}\right\} &=& 0\;,
\label{transport-barq} \\
p^{\mu} {\cal D}_{\mu}G( p,x) + {g \over 2} \: p^{\mu}
\left\{ {\cal F}_{\mu \nu}(x),
{\partial G( p,x) \over \partial p_{\nu}} \right\} &=& 0\;,
\label{transport-gluon}
\end{eqnarray}
\end{mathletters}
$\!\!$where  $g$ is the QCD coupling constant, $\{...,...\}$ denotes the
anticommutator; the covariant derivatives $D_{\mu}$ and ${\cal D}_{\mu}$
act as
$$
D_{\mu} = \partial_{\mu} - ig[A_{\mu}(x),...\; ]\;,\;\;\;\;\;\;\;
{\cal D}_{\mu} = \partial_{\mu} - ig[{\cal A}_{\mu}(x),...\;]\;,
$$
$A_{\mu }$ and ${\cal A}_{\mu }$ being four-potentials
in the fundamental and adjoint representations, respectively:
$$
A^{\mu }(x) = A^{\mu }_a (x) \tau_a \;,\;\;\;\;\;
{\cal A}^{\mu }_{ab}(x) = - if_{abc}A^{\mu }_c (x) \; ,
$$
and  $f_{abc}$ are the structure constants of the $SU(N_c)$ group.
Since the generators of $SU(N_c)$ in the adjoint
representation are given by $(T_a)_{bc} = - i f_{abc}$, one
can also write ${\cal A}^\mu = A^\mu _a T^a$. The stress tensor in
the fundamental representation is $F_{\mu
\nu}=\partial_{\mu}A_{\nu} - \partial_{\nu}A_{\mu} -ig
[A_{\mu},A_{\nu}]$, while  ${\cal F}_{\mu \nu}$ denotes the
field strength tensor in the adjoint representation.

Sometimes it is convenient to project the matrix equations (\ref{transport})
into their colorless and colored components. For the quark distribution
function we define
\be
\label{projec-component}
Q(p,x)   =  {\tilde q}(p,x) + q^a(p,x) \tau^a \; .
\ee
Then, we can deduce from \eq\nr{transport-q} a set of coupled
equations for the different components of $Q$ defined by
\eq\nr{projec-component}. More precisely, we find
\begin{mathletters}
\label{projec-transport}
\ba
 p^\mu \partial_\mu {\tilde q}(p,x) &+& \frac{g}{2 N_c }\, p^\mu
F_{\mu \nu}^a(x) \frac{\partial q^a(p,x)}{\partial p_\nu}  =  0 \; , \\
p^\mu  D_\mu ^{ab} q^b (p,x) & + & \frac{g}{2}\, d^{abc} F_{\mu
\nu}^b(x) \frac{\partial q^c(p,x)}{\partial p_\nu} + g\, p^\mu
F_{\mu \nu}^a(x) \frac{\partial {\tilde q} (p,x)}{\partial p_\nu}
=  0 \; , \ea
\end{mathletters}
$\!\!$where $d^{abc}$ are the totally symmetric structure
constants of $SU(N_c)$ and $D_\mu^{ac} = \partial_\mu \delta^{ac}
+ g f^{abc} A_\mu ^b$. Similar equations can be written for the
antiquark and gluon distribution functions.
 Eqs.~(\ref{projec-transport}) reflect in a very clear way that
transport phenomena of colorless and colored fluctuations are
coupled beyond the lowest order in the gauge coupling constant.
Eqs. (\ref{projec-transport}) might be very useful when collisions
are also taken into account. In this study, however, we find more
convenient to work with the matrix equations (\ref{transport}).

Once the solution of the transport equations is known, we
can obtain the color current associated to the plasma constituents.
The color current is expressed in the fundamental representation as
\be
\label{current}
j^{\mu }(x) = -\frac{g}{2} \int dP \; p^{\mu
} \Big[ Q( p,x) - \bar Q ( p,x) - {1 \over N_c}{\rm Tr}\big[Q(
p,x) - \bar Q ( p,x)\big] +  2i \tau_a f_{abc} G_{bc}( p,x)\Big]
\; ,
\ee
so that $j^\mu_a (x) = 2 {\rm Tr} (\tau_a j^\mu (x))$, and the momentum
measure
\be
dP = \frac{d^4p}{(2 \pi)^3} 2\Theta(p_0)\, \delta(p^2)
\ee
takes into account the mass-shell condition  $p_0 = |{\bf p}|$.
Throughout the paper, we neglect the quark masses, although those
might easily be taken into account by modifying the mass-shell
constraint in the momentum measure. A sum over helicities, two per
particle, and over quark flavors $N_f$ is understood in
\eq\nr{current}, even though it is not explicitly written down.

In the transport theory framework one can consider two different physical
situations: 1) the gauge fields entering into the transport equations
(\ref{transport}) are external, not due to the plasma constituents;
2) the gauge fields can be generated self-consistently by the
quarks and gluons. In the latter case, one also has to solve the
Yang-Mills equation
\be
\label{yang-mills}
D_{\mu} F^{\mu \nu}(x) = j^{\nu}(x)\; ,
\ee
where the color current is given by \eq\nr{current}.

The color current can be derived from an effective action added to
the Yang-Mills one. By means of the relation
\be
\label{current-action}
 j_a^\mu = - {\delta S \over \delta A^a_\mu}
\;, \ee
 where $S \equiv \int d^4x \, {\cal L}\;$, one can obtain
the effective action, up to an integration constant, from the
knowledge of the color current. In the remaining part of this article
we will use this approach to obtain the effective action in different
physical situations.


\section{Exact Solutions}
\label{sec-exact}

Finding exact solutions of the transport equations (\ref{transport}) is in
general a difficult task. However, it is possible to find such solutions
under some restrictive conditions. Here we consider a system where
both the vector gauge field and the distribution functions are
invariant under the space-time translation(s), i.e.
\be
\label{condition1}
\partial_{\alpha_i} A^\mu (x) = 0 \ , \qquad  \mu=0,1,2,3   \ ,
\ee
and
\be\label{homo}
\partial_{\alpha_i}  Q( p,x) =
\partial_{\alpha_i}  \bar Q( p,x) =
\partial_{\alpha_i}  G( p,x) = 0 \;,
\ee
for a fixed $\alpha_i$, where $\alpha_i$ can involve more than one
Lorentz index. For example, if $\alpha_i = 0$ the system is static
while for $\alpha_i =1,2,3$ the gauge field and the distribution
functions depend only on time. The condition Eq.~(\ref{condition1})
is a choice of gauge that, as we will show below, allows one to find
solutions to the transport equations that respect the translation
invariance of the system.

To solve Eqs.~(\ref{transport}) for translation invariant
systems along the direction $x^{\alpha_i}$, we will take into
account two known facts. First, in an electromagnetic plasma, an
exact solution of the corresponding transport equation is given by
any function of the canonical momentum $p_{\alpha_i} -
eA_{\alpha_i}(x)$, if $\partial_{\alpha_i} A_\mu =0$ \cite{hakim}.
Second,  the non-Abelian transport equations of particles carrying
a classical color charge $I^a$ for these translation invariant
systems are also solved by any function of the canonical momentum
$p_{\alpha_i} - g A_{\alpha_i}^a (x) I^a$, if $\partial_{\alpha_i}
A_\mu ^a = 0$ \cite{Laine:2001my}. Because the first case
represents the Abelian limit of the transport equations
(\ref{transport}), while the second one corresponds to the limit of
high-dimensional color representations\footnote{While we are
considering here the transport equations only for particles
carrying color in the fundamental and adjoint representations, the
equations are expected to have the same structure in any other
non-Abelian representation.}, one expects that the solutions of
Eqs.~(\ref{transport}) are of the form:
\begin{mathletters}
\label{ansatz}
\ba
\label{ansatz-q}
 Q( p,x) &=& f(p_{\alpha_i} - gA_{\alpha_i}(x) ) =
\sum_{n=0}^\infty \frac{(-g)^n}{n!} A_{\alpha_1}(x)\,
A_{\alpha_2}(x) \cdots A_{\alpha_n}(x)\, \frac{\partial^n
f(p_{\alpha_i})} {\partial p_{\alpha_1}\, \partial p_{\alpha_2}
\ldots \, \partial p_{\alpha_n}} \; ,
\\
\bar Q( p,x) &=& \bar f(p_{\alpha_i} + gA_{\alpha_i}(x) ) =
\sum_{n=0}^\infty \frac{g^n}{n!} A_{\alpha_1}(x)\, A_{\alpha_2}(x)
\cdots A_{\alpha_n}(x)\, \frac{\partial^n \bar f(p_{\alpha_i})}
{\partial p_{\alpha_1}\, \partial p_{\alpha_2} \ldots \partial
p_{\alpha_n}} \;,
\\
G( p,x) &=& f_g(p_{\alpha_i} - g{\cal A}_{\alpha_i}(x) ) =
\sum_{n=0}^\infty \frac{(-g)^n}{n!} {\cal A}_{\alpha_1}(x)\, {\cal
A}_{\alpha_2}(x) \cdots {\cal A}_{\alpha_n}(x)\, \frac{\partial^n
f_g(p_{\alpha_i})} {\partial p_{\alpha_1}\,
\partial p_{\alpha_2} \ldots \partial p_{\alpha_n}} \;,
\label{ansatz-gluon} \ea
\end{mathletters}
$\!\!$where it is understood that a sum is taken over the
repeated indices. The functions $f$, $\bar f$ and $f_g$ are,
in principle, arbitrary but they can be fixed by additional
considerations.

Let us note that  the distribution functions given by
Eqs.~(\ref{ansatz}) transform covariantly, i.e. according to
Eqs.~(\ref{Q-transform})  and (\ref{G-transform}),
even though the potentials $A^{\mu}$, ${\cal A}^{\mu}$, in general, do not.
Indeed,
\be\label{A-transform}
A^{\alpha_i}(x) \rightarrow M(x)A^{\alpha_i}(x)M^{\dag }(x)
-{i \over g} \big(\partial^{\alpha_i}M(x) \big) M^{\dag }(x) \;.
\ee
However, the second term in the r.h.s of Eq.~(\ref{A-transform}), which
transforms non-covariantly, is eliminated because of the condition
(\ref{condition1}).

Let us check under which conditions the ansatz
\eq\nr{ansatz} solves the transport equations. We first note
that \eq\nr{ansatz-q} is totally symmetric under the exchange
of indices $\alpha_1, \cdots, \alpha_n$. Inserting
\eq\nr{ansatz-q} into the transport equation
(\ref{transport-q}),  one finds
\be
 \label{drift}
p^{\mu} D_{\mu}Q( p,x) = p^{\mu} \sum_{n=0}^\infty
\frac{(-g)^n}{n!} \sum_{s=0}^{n-1} A_{\alpha_1} \cdots
A_{\alpha_s}
 (D_{\mu} A_{\alpha_{s+1}}) \cdots  A_{\alpha_{n}}
\frac{\partial^n f(p_{\alpha_i})} {\partial p_{\alpha_1}\,
\partial p_{\alpha_2} \ldots \partial p_{\alpha_n}} \; , \ee
\be \label{vlasov} {g \over 2}\: p^{\mu}\left\{ F_{\mu \nu}(x),
{\partial Q( p,x) \over \partial p_\nu}\right\} = - p^{\mu}
\sum_{n=0}^\infty \frac{(-g)^{n+1}}{2\, n!} \Big\{D_{\mu}
A_{\alpha_i}, A_{\alpha_1} \cdots   A_{\alpha_n} \Big\}
\frac{\partial^{n+1} f(p_{\alpha_i})} {\partial p_{\alpha_i}\,
\partial p_{\alpha_1} \ldots \, \partial p_{\alpha_n}} \; ,
 \ee where
we have used the following property of a commutator
$$
[X, Y^n] = \sum_{s=0}^{n-1} Y^s [X,Y] Y^{n-s-1} \;
$$
to derive Eq.~(\ref{drift}). We have also taken into account that
$F_{\mu \alpha_i} = D_\mu A_{\alpha_i}$ because of
\eq\nr{condition1}.

It is easy to observe that the terms in \eq\nr{drift} and
\eq\nr{vlasov} that correspond to the same order of the derivative
of $f$ cancel each other exactly if
 \be \label{condition2}
 [D_{\mu}A_{\alpha_i}, A_{\alpha_j}] =0 \ , \qquad
\mu = 0,1,2,3 \; ,
\ee
where $A_{\alpha_j}$ is also in the direction of the translation
invariance.
The same condition is obtained for the antiquark distribution
function, while for gluons one finds that \eq\nr{ansatz-gluon}
is an exact solution of \eq\nr{transport-gluon} if
\be \label{condition2-g}
[{\cal D}_{\mu}{\cal A}_{\alpha_i}, {\cal A}_{\alpha_j}] =0
 \ , \qquad
\mu = 0,1,2,3 \ .
\ee
However, it is easy to prove that \eq\nr{condition2-g} is
automatically satisfied if \eq\nr{condition2} holds.

For static systems ($\alpha=0$), \eq\nr{condition2} reduces to the
commutation relation between the color electric field and $A_0$.
In a more general situation, the condition \eq\nr{condition2}
simplifies the non-Abelian field dynamics in the direction
of the translation invariance. Note that the commutation condition
is trivially satisfied in the Abelian limit. If $T_R^a$ is a generator
of a representation $R$ of $SU(N_c)$,  then, after a normalization of
these generators, one would get  $[T_R^a,T^b_R] \rightarrow 0$ for
high-dimensional representations. Consequently, the commutation
condition would also be satisfied. This explains how to reconcile
the matrix results with those obtained with the non-Abelian transport
equations for classical color.

Once the solution for the quark, antiquark and gluon distribution
functions are known, one can compute the color current.
Inserting (\ref{ansatz}) into Eq.~(\ref{current}), we get
\ba \label{current2}
 j_a^{\mu}(x) &=& 2 \sum_{n=0}^\infty
\frac{(-g)^{n+1}}{n!} A_{\alpha_1}^{c_1} (x) \cdots
A_{\alpha_n}^{c_n} (x)
 \int dP \;  p^\mu \Bigg[N_f {\rm Tr} [\tau_a \tau_{c_1} \cdots \tau_{c_n}]
\bigg[\frac{\partial^n f(p_{\alpha_i})} {\partial p_{\alpha_1}
\cdots \partial p_{\alpha_n}}
\\ \nonumber
&+&  (-1)^{n+1} \frac{\partial^n {\bar f}(p_{\alpha_i})} {\partial
p_{\alpha_1} \cdots \partial p_{\alpha_n}} \bigg]+ {\rm Tr} [T_a
T_{c_1} \cdots T_{c_n}] \frac{\partial^n  f_g
(p_{\alpha_i})}{\partial p_{\alpha_1} \cdots
\partial p_{\alpha_n}}
\Bigg]  \; .
\ea
The factor $2$ in the above equation arises from the two
helicities associated with every particle species. When the
functions $f$, $\bar f$ and $f_g$ are determined, one can evaluate
the momentum integral of \eq\nr{current2}, and then, after
solving \eq\nr{current-action}, one obtains the associated
effective action.


\section{Perturbative solutions}
\label{pert-sol}

\subsection{General Considerations}
\label{pert-general}

In \se \ref{sec-exact} we have found exact solutions of the
collisionless transport equations (\ref{transport}) for translation
invariant systems that obey the extra condition \eq\nr{condition2}.
In this section we treat the transport equations perturbatively
and find solutions that are not constrained by the condition
\eq\nr{condition2}.

We assume here that
\be g \ll 1 \;, \ee
i.e. we deal with the weak coupling regime of the theory,
and consider an expansion of the distribution function of the
form
\be \label{expansion}
Q = Q^{(0)}+ Q^{(1)}+ Q^{(2)}+ ....
\ee
where the  $0$-th term is a known function
\be
\label{unpert}
Q^{(0)}( p,x) = f(p_{\alpha_i}) \;.
\ee
The higher-order terms are determined by the equation
\be
\label{iteration} p^{\mu} D_{\mu}Q^{(n)}( p,x) + {g \over 2}\:
p^{\mu} \left\{ F_{\mu \nu}(x), {\partial Q^{(n-1)}( p,x) \over
\partial p_{\nu}}\right\} =  0\;.
\ee
Of course, the same treatments should be followed to study the
antiquark and gluon distribution functions, but those are nearly
identical.

The iterative procedure based on \eq\nr{iteration} does not
correspond to a strict expansion in powers of $g$. Such an
expansion would force us to split all covariant derivatives
into a derivative and a commutator part, resulting in the
breaking of the gauge covariance of every term in the perturbative series.
We maintain the gauge covariance of every term in \eq\nr{expansion}
at the expense of reorganizing the perturbative expansion.

It should be noticed that for $f$ being the equilibrium distribution 
function, the first term ($n=1$) of the above perturbation series reproduces 
the hard thermal loops of QCD \cite{Braaten:1989mz,Bla93}. Let us also stress 
that while we are applying here a perturbative method to the
transport equations in their matrix form, fully equivalent
results can be obtained using the projected equations
(\ref{projec-transport}). Such an approach has been carried
out by B\"odeker and Laine to order $g^2$ for static
quasi-equilibrium systems \cite{Bod01}. We extend
that analysis by pushing the perturbative procedure to higher
orders in $g$. We do not attempt to solve the transport
equations in full generality, as  the solutions at
every order then turn out to be highly non-local. We reduce
our study to translation invariant systems when the solutions
are much simplified.


\subsection{From $g$ to $g^4$ order}
\label{pert-transl}


In this subsection, we consider a translation invariant system that obeys
\eq\nr{condition1} for a fixed $\alpha_i$. We also assume that the unperturbed
distribution function depends only on $p_{\alpha_i}$ as in \eq\nr{unpert}.
The first-order correction to $Q^{(0)}$ is obtained by solving the equation
\be
p^{\mu} D_{\mu}Q^{(1)} =
 - g \; p^{\mu} F_{\mu \alpha_1}
 \;\frac{\partial f(p_{\alpha_i})}{\partial p_{\alpha_1}} = - g
p^\mu D_{\mu} A_{\alpha_1} \;\frac{\partial
f(p_{\alpha_i})}{\partial p_{\alpha_1}} \;.
\ee
The solution  reads (up to a function $h$, such that $p^\mu D_\mu h =0$,
that we will neglect throughout)
\be Q^{(1)}(p,x) = - g \;A_{\alpha_1}(x) \; \frac{ \partial
f(p_{\alpha_i})}{\partial p_{\alpha_1}} \;,
\ee
and it coincides with the first term of the ansatz (\ref{ansatz-q}).

The transport equation at second-order is \be p^{\mu}
D_{\mu}Q^{(2)} = {g^2 \over 2} \;  \Big \{  p^{\mu} D_{\mu}
A_{\alpha_1}, A_{\alpha_2}  \Big\}
\;\frac{\partial^2f(p_{\alpha_i})}{\partial p_{\alpha_1}
\partial p_{\alpha_2}} \;,
\ee
but it can be rewritten as
\be
p^{\mu} D_{\mu}Q^{(2)} = {g^2 \over 2} \;
p^{\mu} D_{\mu}\left( A_{\alpha_1} A_{\alpha_2} \right)
 \frac{\partial^2 f(p_{\alpha_i})}{\partial p_{\alpha_1}
\partial p_{\alpha_2}}
 \;.
\ee
Thus, the second-order solution is
\be
Q^{(2)}(p,x) = {g^2 \over 2} \;
 A_{\alpha_1}(x) A_{\alpha_2}(x)
\; \frac{\partial^2 f(p_{\alpha_i})}{\partial p_{\alpha_1}
\partial p_{\alpha_2}}
 \;,
\ee
which again coincides with the respective term of the ansatz
(\ref{ansatz-q}).

The third-order equation is
\be
\label{boltzmann3} p^{\mu}
D_{\mu}Q^{(3)} = - {g^3 \over 4} \; \Big\{ p^{\mu} D_{\mu}
A_{\alpha_1}, A_{\alpha_2} A_{\alpha_3} \Big \} \;\frac{\partial^3
f(p_{\alpha_i})}{\partial p_{\alpha_1} \partial p_{\alpha_2}
\partial p_{\alpha_3}} \;,
\ee
and its r.h.s. is {\it not}
proportional to a covariant derivative of a third power of
$A_{\alpha_i}$. However, the  anticommutator of the r.h.s.
of the equation can be rewritten as follows:
\be
p^\mu D_\mu Q^{(3)}  =
-\frac{g^3}{3!} \left(
p^{\mu} D_{\mu}
\left( A_{\alpha_1} A_{\alpha_2} A_{\alpha_3} \right) + \frac12
\Big[ [p^{\mu} D_{\mu} A_{\alpha_1},A_{\alpha_2}],A_{\alpha_3}
\Big] \right) \frac{\partial^3 f(p_{\alpha_i})}{\partial
p_{\alpha_1} \partial p_{\alpha_2}
 \partial p_{\alpha_3}} \ .
\ee
The term with the commutator cannot be expressed as a total
covariant derivative. Thus, the solution of the above equation
contains two pieces: a term that is local in the gauge fields
and a non-local part. Namely,
\be
\label{nlboltzmann3}
Q^{(3)} (x,p)  =
-\frac{g^3}{3!} \, A_{\alpha_1}(x) A_{\alpha_2}(x) A_{\alpha_3}(x) \,
\frac{\partial^3 f(p_{\alpha_i})}{\partial
p_{\alpha_1} \partial p_{\alpha_2}
 \partial p_{\alpha_3}} +
Q^{(3)}_{\rm nl}(x,p) \;
\ee
with the non-local term obeying the equation
\be
\label{nlboltn3}
 p^\mu D_\mu
Q^{(3)}_{\rm nl}  = -\frac{g^3}{12}
\Big[ [p^\mu D_\mu A_{\alpha_1},A_{\alpha_2}],A_{\alpha_3} \Big]
\frac{\partial^3 f(p_{\alpha_i})}{\partial
p_{\alpha_1} \partial p_{\alpha_2}
\partial p_{\alpha_3}}  \;.
\ee
One observes that when the condition \eq\nr{condition2} is
satisfied, $Q^{(3)}_{\rm nl} = 0$ and we recover the solution
of  order $g^3$ of \eq\nr{ansatz-q}. But this term is non-zero
under more general circumstances. We also notice that if
\be
[\partial_\mu A_{\alpha_i}, A_{\alpha_j}] = 0 \;,
\ee
the non-local piece is  proportional to $g^4$.

Let us solve \eq\nr{nlboltn3}. Using the $SU(N_c)$ algebra, we
rewrite the commutator as
\be
p^\mu D_\mu Q^{(3)}_{\rm nl}  =
-\frac{g^3}{12} f^{acd} f^{deb} \tau^a
(p^\mu D_\mu A_{\alpha_1})^e
A_{\alpha_2}^c A_{\alpha_3}^b
\frac{\partial^3 f(p_{\alpha_i})}{\partial
p_{\alpha_1} \partial p_{\alpha_2}
\partial p_{\alpha_3}} \equiv R^a \,\tau^a \ .
\ee
Consequently, $Q^{(3)}_{\rm nl} = Q^{(3) a}_{\rm nl} \tau^a$ with
\be
\label{diff-eq}
p^\mu D_\mu^{ac} Q_{\rm nl}^{(3) c} = R^a \ .
\ee
The solution can be expressed as
\be \label{nl-3boltzmann}
 Q_{\rm nl}^{(3) a} (p,x) = \int d^4 y \,
\langle x| \frac{1}{ p\cdot D}  | y \rangle_{a b} \, R^b (y) \ ,
\ee
where $1/p \cdot D$ is the retarded Green function
associated to the differential equation (\ref{diff-eq}).
The explicit form of $1/p \cdot D$ can be found, for example, in
Sec. II of \cite{Mro00}.

The fourth-order equation reads
\be
\label{boltzmann4}
p^{\mu} D_{\mu}Q^{(4)} =
\frac{g^4}{12} \Big\{ p^{\mu} D_{\mu} A_{\alpha_1},
A_{\alpha_2} A_{\alpha_3} A_{\alpha_4}\Big\} \;\frac{\partial^4
f(p_{\alpha_i})}{\partial p_{\alpha_1} \partial p_{\alpha_2}
\partial p_{\alpha_3} \partial p_{\alpha_4}}
 +  \frac{g^4}{24} \Big\{ F_{\mu \nu}, {\partial Q_{\rm
nl}^{(3)} \over
\partial p_{\nu}} \Big\} \ .
\ee
The first anticommutator in the r.h.s. of the equation can also be
rewritten as a total covariant derivative plus an additional
commutator. Thus,
\ba
\nonumber
 p^{\mu} D_{\mu}Q^{(4)} =
\frac{g^4}{4!} \Bigg(  p^{\mu} D_{\mu} \left( A_{\alpha_1}
A_{\alpha_2} A_{\alpha_3} A_{\alpha_4} \right) &+& \Big[ [p^{\mu}
D_{\mu} A_{\alpha_1},A_{\alpha_2}],A_{\alpha_3} A_{\alpha_4}
\Big] \Bigg) \frac{\partial^4 f(p_{\alpha_i})}{\partial
p_{\alpha_1} \partial p_{\alpha_2}
 \partial p_{\alpha_3} \partial p_{\alpha_4}} \\
& + & \frac{g^4}{4!} \Big\{ F_{\mu \nu}, {\partial Q_{\rm
nl}^{(3)} \over
\partial p_{\nu}} \Big\} \ .
\ea
We see that the solution is of the form
\be
\label{nlboltzmann4}
Q^{(4)}(x,p)  =
\frac{g^4}{4!} \, A_{\alpha_1}(x) A_{\alpha_2}(x) A_{\alpha_3}(x)
A_{\alpha_4} (x) \,
\frac{\partial^4 f(p_{\alpha_i})}{\partial
p_{\alpha_1} \partial p_{\alpha_2}
 \partial p_{\alpha_3} \partial p_{\alpha_4}} +
Q^{(4)}_{\rm nl}(x,p) \;,
\ee
where $Q^{(4)}_{\rm nl}$ represents the non-local contribution that
vanishes if the condition \eq\nr{condition2} is satisfied. We note that
the local piece coincides with the term $g^4$ of the ansatz (\ref{ansatz-q}).

In principle, we could solve the equation for $Q^{(4)}_{\rm nl}$ and
for the following terms of the perturbative expansion, finding that every
term contains both local and non-local pieces in the gauge fields.
However, we stop our analysis here as the structure of the non-local
terms becomes more and more complex.

Adding the contributions of quarks, antiquarks and gluons, we can
obtain the color current. Up to the term $n=3$, the color current
still has a relatively simple form and is given by
\ba
 \label{pert-current}
 j_a^{\mu}(x) &=& 2 \sum_{n=0}^3
\frac{(-g)^{n+1}}{n!} A_{\alpha_1}^{c_1} (x) \cdots
A_{\alpha_n}^{c_n} (x)
 \int dP \;  p^\mu \Bigg[N_f {\rm Tr} [\tau_a \tau_{c_1} \cdots \tau_{c_n}]
\bigg[\frac{\partial^n f(p_{\alpha_i})} {\partial p_{\alpha_1}
\cdots \partial p_{\alpha_n}}
\\ \nonumber
&+&  (-1)^{n+1} \frac{\partial^n {\bar f}(p_{\alpha_i})} {\partial
p_{\alpha_1} \cdots \partial p_{\alpha_n}} \bigg]+ {\rm Tr} [T_a
T_{c_1} \cdots T_{c_n}] \frac{\partial^n  f_g
(p_{\alpha_i})}{\partial p_{\alpha_1} \cdots
\partial p_{\alpha_n}}
\Bigg]  + j^\mu_{a, {\rm nl}} (x) \;,
\ea
 where the non-local term reads
\ba
\label{non-localcurrent}
 j^\mu_{a ,{\rm nl}} (x)  & = &  \frac{g^4}{6}  \int dP  \int
d^4 y \, \langle x| \frac{p^\mu}{ p\cdot D}  | y
\rangle_{ab}\Bigg[ 2 N_f {\rm Tr} \left(\tau^b \bigg[[ p \cdot D
A_{\alpha_1}, A_{\alpha_2}], A_{\alpha_3} ] \bigg] \right) \bigg[
 \frac{\partial^3 f(p_{\alpha_i}) }{\partial p_{\alpha_1} \partial
p_{\alpha_2}
 \partial p_{\alpha_3}}
\nonumber \\
& + &  \frac{\partial^3 {\bar f}(p_{\alpha_i}) }{\partial
p_{\alpha_1} \partial p_{\alpha_2}
 \partial p_{\alpha_3}}   \bigg]+ \frac{1}{N_c} {\rm Tr} \left(T^b
 \bigg[[ p \cdot {\cal D} {\cal A}_{\alpha_1}, {\cal
A}_{\alpha_2}], {\cal A}_{\alpha_3} ] \bigg] \right)
\frac{\partial^3 f_g(p_{\alpha_i}) }{\partial p_{\alpha_1}
\partial p_{\alpha_2}
 \partial p_{\alpha_3}} \Bigg] \;.
\ea

In the following section we will get the effective action for
static systems close to equilibrium. In that case, it is easy
to see that for $n \geq 4$ the local pieces do not contribute
to the current or effective action, as either the momentum
integrals (Appendix \ref{integrals}) or the traces of generators
in the adjoint representation (Appendix \ref{traces}) vanish.
However, the same does not hold true for the non-local terms.


\section{Effective action of quasi-equilibrium systems}
\label{actions}

In this section we find the effective action for  systems close
to thermal equilibrium. Then, the functions  $f$, $\bar f$ and
$f_g$ are of the Fermi-Dirac or Bose-Einstein form
\be \label{eq-distribution}
f_{\rm FD}(E) = {1\over e^{\beta(E-\mu)} + 1} \;,\;\;\;\;\;\;\;
\bar f_{\rm FD}(E) = {1 \over e^{\beta(E+\mu)} + 1} \;,\;\;\;\;\;\;\;
f_{\rm BE}(E)= {1\over e^{\beta E} - 1} \; ,
\ee
where $\beta \equiv 1/T$, and $T$ is the temperature, and
$\mu$ is the quark chemical potential.

\subsection{Static Systems}

We first consider  static systems satisfying the condition
\be
\label{static-condition}
 [D_\mu A_0, A_0 ] = 0 \ , \qquad \mu=0,1,2,3 \;.
\ee
The color current is given by \eq\nr{current2} with
$\alpha_i=0$ while the distribution functions are those
in Eq.~(\ref{eq-distribution}).

After performing the momentum integral, one observes that
$j_a^i = 0$ since
\be \int
\frac{d\Omega_{\bf v}} {4 \pi} \, v^i = 0 \ ,
\ee
where $v^i = p^i/|{\bf p}|$ is the particle velocity, and the
integral is performed over angular directions of ${\bf v}$.
Thus, only $j_a^0$, i.e. the color current density, is non-vanishing.
In the case of static systems, one easily finds the Lagrangian density
by integrating \eq\nr{current-action}. The fermionic
contribution arises from
\be
 \label{lagrangian1-quark} {\cal L}_f = - { N_f \over
\pi^2} \sum_{n=0}^\infty \frac{(-g)^{n+1}}{(n+1)!} \int_{\Lambda}
^{\infty}
dE \,E^2 \;
 \bigg[\frac{d^nf_{\rm FD}(E)}{dE^n} + (-1)^{n+1} \frac{d^n \bar
f_{\rm FD}(E)}{dE^n} \bigg] \; {\rm Tr}[A_0^{n+1}(\bfx)] \;, \ee
while that of gluons is
\be \label{lagrangian1-gluon}
{\cal L}_g
= - { 1\over \pi^2} \sum_{n=0}^\infty \frac{(-g)^{n+1}}{(n+1)!}
\int_{\Lambda}^{\infty} dE \,E^2 \; \frac{d^n f_{\rm BE}(E)}{dE^n}  \; {\rm
Tr} [{\cal A}_0^{n+1}(\bfx)] \;.
\ee
In the above expressions we have made use of the mass-shell
condition, which gives $E \equiv p_0 = |{\bf p}|$, and we have put
an infrared cut-off, $T \gg \Lambda \gg gT$, which excludes the
contribution of the soft modes to the integrals, as only the hard
modes behave as quasiparticles.

We note that the above Lagrangians can be written as
\ba
\label{L-f}
 {\cal L}_f &=& {T N_f  \over \pi^2} \; {\rm Tr}
\int_{\Lambda}^{\infty} dE \,E^2 \;
 \Big[ {\rm ln}\big(1+e^{-\beta (E-\mu -gA^0)}\big) + {\rm
ln}\big(1+e^{-\beta (E+\mu + gA^0)} \big) \Big] \ \;,
\\ \label{L-g}
{\cal L}_g &=& {T  \over \pi^2} \; {\rm Tr} \int_{\Lambda}^{\infty} dE
\,E^2 \;  {\rm ln}\big(1- e^{-\beta (E -g{\cal A}^0)} \big)  \;,
\ea
because
$$
{d \over dx} \; {\rm ln}(1\pm e^{-x}) = \pm {1 \over e^x \pm 1}
\;.
$$
The Lagrangians (\ref{lagrangian1-quark}, \ref{lagrangian1-gluon})
differ from those given in Eqs.~(\ref{L-f}, \ref{L-g}) by  field
independent terms corresponding to the free-quark and free-gluon
contributions to the system's pressure.

While it is not obvious from Eq.~(\ref{lagrangian1-gluon}),
Eq.~(\ref{L-g})  shows that the gluon action is {\em not} well
defined without the infrared cut-off because for energies  smaller
than the eigenvalues of $ g A_0$, which are typically of order
$gT$,
 the logarithm
in Eq.~(\ref{L-g}) becomes a multi-valued complex function. Of
course, there is an analogous problem with the Lagrangian
\eq\nr{lagrangian1-gluon}. The difficulty one encounters is not
surprising - the kinetic description breaks down for sufficiently
soft modes. Then, the equilibrium distribution function of gluons
of the form (\ref{ansatz-gluon}) becomes negative, loosing its
probabilistic interpretation.

We leave the evaluation of the energy integrals from
Eqs.\,({\ref{lagrangian1-quark}) and  (\ref{lagrangian1-gluon})
for the Appendix \ref{integrals}, where we first compute the
integrals with the cut-off. However, for all contributions the
cut-off dependent terms appear to be regular and subdominant.
Therefore, we take the limit $\Lambda \rightarrow 0$ and thus
drop all the cut-off dependent terms\footnote{A matching 
procedure with the soft classical field theory allows one to 
eliminate the cut-off dependent terms but we will not pursue 
that procedure here.}. As a result, we find the field dependent 
fermionic contribution to the system's Lagrangian as
\be
 \frac{ {\cal L}_{f}}{N_f} = g
\,\frac{\mu}{3} \Bigl(T^2 + \frac{\mu^2}{\pi^2}\Bigr) {\rm Tr} A_0
+ \frac{g^2}{2} \Bigl(\frac{T^2}{3} + \frac{\mu^2}{\pi^2}\Bigr)
{\rm Tr} A_0^2 + \mu \frac{g^3}{3\pi^2}{\rm Tr} A_0^3 +
\frac{g^4}{12\pi^2} {\rm Tr} A_0^4  \;. \la{LMf}
\ee We have kept
here a linear term in the gauge potential that only survives in
the Abelian limit where ${\rm Tr} A_0 = A_0$. Taking into account
that, as shown in  Appendix \ref{traces}, the symmetric traces of
an odd number of adjoint generators vanish, we find the following
gluon contribution \be {\cal L}_g  = \frac{g^2 T^2}{6}  {\rm Tr}
{\cal A}_0^2  - \frac{g^4}{24\pi^2} {\rm Tr} {\cal A}_0^4  \,.
\la{LMg} \ee It is also shown in Appendix \ref{traces} that the
traces of the fields in the adjoint representation can be
expressed through the fundamental representation traces.

As already stressed, the integral (\ref{L-g}) is ill-defined
without the infrared cut-off. However, the integral with $\Lambda
= 0$ is regular after a Wick rotation to Euclidean space-time. For
constant background fields $A_0$, and after the Wick rotation $A_0
\rightarrow i A^E_0$ we observe that Eqs.~(\ref{L-f}) and 
(\ref{L-g}) with $\Lambda = 0$ {\em fully agree} with the
complete one-loop contribution to the effective potential for the
phase of the Polyakov line \cite{Gross:1980br,Weiss:1980rj,cka}.
It should be mentioned that a constant background field $A_0$ 
can always be chosen in diagonal form \cite{Gross:1980br}, and 
thus satisfies the condition \eq\nr{static-condition}. The Wick 
rotated integral (\ref{L-g}) is discussed in Appendix 
\ref{one-loop}. We note here that except for the terms analogous 
to those present in Eq.~(\ref{LMg}) it provides non-analytic 
terms which are cubic in the eigenvalues of $A^E_0$. We also 
observe that Eqs.~(\ref{LMf}) and (\ref{LMg}) after the Wick 
rotation {\em agree} with the dimensionally reduced effective 
action \cite{Har00}.

Now, let us briefly consider a static system that does not
satisfy \eq\nr{static-condition}. Using the solutions
of \se \ref{pert-transl}, we find that up to the order $g^4$
Eqs.~(\ref{LMf}) and (\ref{LMg}) still hold, but at  order
$g^4$ and beyond there are corrections due to the non-local
terms. The first non-local contribution appears at  order
$g^4$ from the non-local current \eq\nr{non-localcurrent} with
$\alpha_i =0$. To get the corresponding term in the effective
action, one  should still solve \eq\nr{current-action}.
We have not found the solution, but we also see no reason
why the term should vanish. Thus, we conclude that the presence
of the non-local contribution signals a discrepancy with the
dimensionally reduced effective action \cite{Har00} at
order $g^4$, whenever the condition \eq\nr{static-condition}
is not satisfied.


\subsection{Homogeneous Systems}


We consider here  time-dependent homogeneous systems that obey the
condition \eq\nr{condition2} with $\alpha_i =1,2,3$. For a system close to
equilibrium, where the rotational symmetry is not broken, the solution of
the transport equations can only depend on the modulus of the
canonical three-momentum $|{\bf p} - g {\bf A}(t)|$. The color
current in this situation is given by \eq\nr{current2} with
$f(p_{\alpha_i}), \bar f(p_{\alpha_i})$ and $f_g(p_{\alpha_i})$
replaced by $f_{\rm FD} (|\bfp|), \bar f_{\rm FD} (|\bfp|)$,
and $f_{\rm BE} (|\bfp|)$, respectively.

The color current is then given by \eq\nr{current2}, but now $\alpha_i= 1,2,3$.
To get the expression for the color current, we first need to perform
the momentum integral of \eq\nr{current2}. The partial derivatives
appearing in \eq\nr{current2} can be explicitly evaluated. Using
the chain rule, they can be expressed as partial derivatives of $E=|\bfp|$.
Therefore,
$$
\frac{\partial f}{\partial p_i} =
\frac{\partial E}{\partial p_i} \frac{d f}{dE} \ , \qquad
\frac{\partial^2 f}{\partial p_{i} \partial p_{j} } =
\frac{\partial^2 E}{\partial p_{i}
\partial p_{j}} \frac{d f}{dE}
+ \frac{\partial E}{\partial p_{i}}
 \frac{\partial E}{\partial p_{j}} \frac{d^2 f}{dE^2} \;,
$$
\ba
\nonumber
 \frac{\partial^3f}{\partial p_{i} \partial p_{j} \partial p_{k} }
&=& \frac{\partial^3 E}{\partial p_{i}
\partial p_{j} \partial p_{k}} \frac{d f}{dE} +
\left( \frac{\partial E}{\partial p_{i}}
 \frac{\partial^2 E}{\partial p_{j} \partial p_{k}} +
\frac{\partial E}{\partial p_{j}}
 \frac{\partial^2 E}{\partial p_{i} \partial p_{k}} +
\frac{\partial E}{\partial p_{k}}
 \frac{\partial^2 E}{\partial p_{i} \partial p_{j}}
\right) \frac{d^2 f}{dE^2} \\
& + & \frac{\partial E}{\partial p_{i}}
 \frac{\partial E}{\partial p_{j}}
\frac{\partial E}{\partial p_{k}} \frac{d^3 f}{dE^3} \ ,
\qquad {\rm etc} \ , \nonumber
\ea
where $i,j, k = 1,2,3$. The partial derivatives of $E$ can
be written as functions of the particle velocities. Let
us define
\be
L^{i_1 \cdots i_n}_{(n)} \equiv \frac{\partial^n E }{\partial
p_{i_1} \cdots \partial p_{i_n}} \ .
\ee
An explicit evaluation of the first terms gives
$$
L^i_{(1)} = v^i \ , \qquad L^{i j}_{(2)}  = \frac{1}{E}
\left(\delta^{i j} - v^{i} v^{j} \right) \ ,
$$
$$
L^{i j k}_{(3)}  =  - \frac{1}{E} \left( L^{i j}_{(2)}
v^{k} +  L^{i k}_{(2)} v^{j}
 + L^{j k}_{(2)} v^{i} \right) \ ,  \qquad {\rm etc}.
$$

The angular integral of every term in the series of \eq\nr{current2}
can be easily calculated. Since the integrals of an even number of
$v^i$ vanish we only need:
\ba
\int \frac{d
\Omega_{\bf v}} {4 \pi} \, v^{i} v^{j} &  = & \frac 13 \, \delta^{i j}
\ , \\
\int \frac{d \Omega_{\bf v}} {4 \pi} \, v^{i} v^{j} v^{k}
v^{l} &  = & \frac{1}{15} \left( \delta^{i j}\delta^{k
l} + \delta^{i k}\delta^{j l} + \delta^{i
l}\delta^{j k} \right) \;.
\ea
The integrals over $E$ are the same as those used in the previous
subsection (see  Appendix \ref{integrals}).

The evaluation of the momentum integrals gives, as expected,
$j^0_a = 0$. The space components of the color current are,
of course, non-zero with the quark and gluon contributions
given, respectively, as
\ba
j^j_a &=&  \frac {g^2}{3} \Bigl(\frac{T^2}{3}
+ \frac{\mu^2}{\pi^2}\Bigr) {\rm Tr}(\tau_a A^j) \nonumber \\
&+& \frac{g^4}{45 \pi^2} \Bigl(  {\rm Tr} (\tau_a A^j A^i A^i)
+  {\rm Tr} (\tau_a A^i A^j A^i)
+  {\rm Tr} (\tau_a A^i A^i A^j)  \Bigr) \;, \\ [3mm]
j^j_a &=&  \frac {g^2 T^2}{9}  {\rm Tr}(T_a {\cal A})
+ \frac{g^4}{90 \pi^2} \Bigl( {\rm Tr} (T_a {\cal A}^j {\cal A}^i {\cal A}^i)
+   {\rm Tr} (T_a {\cal A}^i {\cal A}^j {\cal A}^i)
+ {\rm Tr} (T_a {\cal A}^i {\cal A}^i {\cal A}^j)  \Bigr) \;.
\ea
These currents arise from the following Lagrangian densities:
\be
\label{Ltq}
\frac{ {\cal L}_{f}}{N_f}  = -
\frac{g^2}{6} \Bigl(\frac{T^2}{3}
+ \frac{\mu^2}{\pi^2}\Bigr) {\rm Tr} (A_j A_j)
- \frac{g^4}{ 90 \pi^2} \left( {\rm Tr} (A_j A_j A_i A_i)
+ \frac{1}{2} {\rm Tr} (A_j A_i A_j A_i) \right)
\ ,
\ee
and
\be
\label{Ltg}
{\cal L}_g   = -
\frac{g^2 T^2}{18}  {\rm Tr} ({\cal A}_j {\cal A}_j)
- \frac{g^4}{180\pi^2}
  \left( {\rm Tr} ({\cal A}_j {\cal A}_j {\cal A}_i {\cal A}_i)
+ \frac 12  {\rm Tr} ({\cal A}_i {\cal A}_j {\cal A}_i {\cal A}_j)
\right) \ .
\ee

We should point out that the terms proportional to $g^2$ in
\eq\nr{Ltq} and \eq\nr{Ltg} agree with the hard thermal loop
Lagrangians in the homogeneous limit \cite{Braaten:1989mz}. If the
condition \eq\nr{condition2} is not satisfied, the color current,
and thus the associated effective Lagrangians, are corrected at
order $g^4$ and beyond by the addition of the non-local terms,
exactly as happened for the static systems. However, we will not
write down these explicit terms here.


\section{Conclusions}
\label{discussion}


Our results show  the efficiency of transport theory in
describing the quark-gluon plasma at soft scales.
We have shown how the solutions of the collisionless transport
equations for the static systems close to equilibrium reproduce
the one-loop effective potential for the phase of the Polyakov
line. Up to now, the results of the transport theory and quantum
field theoretical computations have been known to agree only for the
lower-dimensional operators, but our computations indicate that the
agreement extends to the full one-loop effective action. We find a complete
match of the transport results with those of the one-loop effective
potential in the presence of a constant background field.
For non-constant static background fields, transport theory
predicts the appearance of non-local operators in the effective
action, starting at order $g^4$ and beyond. This result then suggests
a discrepancy with the dimensionally reduced effective theories \cite{Har00},
where these non-local operators are not present.

We have limited our analysis to translation invariant systems,
when the solutions of the transport equations are local in the
lower orders of the perturbative expansion. It would be
desirable to solve the equations in full generality. However,
the solutions are then complex non-local functions of the gauge
fields, and their structure beyond order $g^2$ is not particularly
enlightening. It is presumably more promising to explore the combined
set of equations (\ref{transport}) and (\ref{yang-mills}) numerically,
thus allowing for a non-perturbative study of dynamical phenomena
at soft scales, even beyond the hard thermal loop approximation.

\acknowledgements

We wish to thank M. Laine for his critical reading of the 
manuscript and fruitful discussions. We are also indebted to 
J.L.F. Barb\'on, M. Garc\'{\i}a-Perez,  K. Korthals Altes  
and O. Philipsen for useful conversations. C. M. was partially 
supported by the EU through Contract No. HPMF-CT-1999-00391. 
Our special thanks go to the ITP at Santa Barbara where the 
project was initiated during the Workshop `QCD and Gauge Theory 
Dynamics in the RHIC Era'. We are grateful to the NSF for 
a support under Grant No. PHY99-07949.

\appendix


\section{Momentum Integrals}
\label{integrals}



\subsection{Bosons}


The bosonic integrals to be evaluated are of the form:
\be
\label{bosonic1}
 I_n^b (\Delta)= T^{3-n} \int_{\Delta}^\infty dx \,x^2 \;
\frac{d^nf_{\rm BE}(x)}{dx^n} \;,
\ee
where $\Delta \equiv \Lambda/T$. Expanding the Bose-Einstein
distribution as
$$
f_{\rm BE} (x) = {1 \over e^x -1}=\sum_{m=1}^\infty e^{-m x}
$$
and interchanging the order of summation and integration,
the integral that has to performed reduces to
\be
\label{integral}
\int_{\Delta}^\infty dx\, x^2 e^{-mx} = e^{-m \Delta} \left(
\frac{2}{m^3} + \frac{2 \Delta}{m^2} +\frac{\Delta^2}{m} \right) \;.
\ee
Therefore,
\be
\label{bosonic2}
I_n^b (\Delta) = 2 (-1)^n T^{3-n} \left({\rm Li}_{3-n}(e^{-\Delta})
+ \Delta {\rm Li}_{2-n}(e^{-\Delta}) + {\Delta^2 \over 2}
{\rm Li}_{1-n}(e^{-\Delta}) \right)
\ee
where the solution is expressed in terms of the Euler polylogarithm 
function
\be
\label{poly}
{\rm Li}_s(z)  \buildrel \rm def \over =
\sum_{m=1}^\infty {z^m \over m^s} \;.
\ee

The function ${\rm Li}_N(e^{-\Delta})$ can be expanded in powers
of $\Delta$ as \cite{Hab82,haw}
\be \label{Li-sum1}
{\rm Li}_N (e^{-\Delta}) = \sum_{k=0}^\infty (-1)^k
\frac{\Delta^k}{k!}
\zeta(N-k)
\ee
for $N<1$ and
\be \label{Li-sum2}
{\rm Li}_N (e^{-\Delta}) = \sum_{k=0,\; k \neq N-1}^\infty (-1)^k
\frac{\Delta^k}{k!}
\zeta(N-k) + (-1)^N \frac{\Delta^{N-1}}{(N-1)!}
\left(\ln \Delta - H_{N-1} \right)
\ee
for $N>1$; the zeta function is defined as
\be \label{zeta}
\zeta (s)  \buildrel \rm def \over =
\sum_{m=1}^\infty {1 \over m^s} \;, \qquad {\rm Re}s > 1\;.
\ee
and $H_N \equiv 1 + \frac{1}{2} + \frac{1}{3} + \cdots + \frac{1}{N}$.

While the series (\ref{poly}) is convergent for $|z| <1$, the series
(\ref{Li-sum1}, \ref{Li-sum2}) with the zeta function defined by
Eq.~(\ref{zeta}) seem to be divergent because the zeta argument
is repeatedly equal or smaller than 1. This happens due to the `illegal'
interchanging of the two summations. As is well known, the problem
is resolved by means of the zeta function regularization procedure
\cite{Actor:zf,haw}, where the analytic continuation of $\zeta (s)$
instead of the definition (\ref{zeta}) is used. Then, $\zeta (1)$
remains truly divergent while
\ba
\zeta (0) &=& - {1 \over 2} \ ,\nonumber \\
\zeta (1-2k) &=& -{B_{2k} \over 2k} \ , \;\;\;\; k=1,2,3,...
\nonumber \\ \nonumber
\zeta (-2k) &=& 0 \ , \;\;\;\;\;\;\;\;\;\;\; k=1,2,3,...
\ea
where $B_{l}$ are the Bernoulli polynomials.

Using Eqs.~(\ref{Li-sum1}, \ref{Li-sum2}) one finds that when
$\Delta \rightarrow 0$ \be {\rm Li}_N (e^{-\Delta}) = \zeta(N) +
{\cal O}(\Delta ) \ee for $N \not= 1$. Since ${\rm Li}_1 (z) =
-{\rm ln}(1-z)$, ${\rm Li}_1 (e^{-\Delta})$ diverges as $-{\rm
ln}\Delta$. Therefore, when $\Delta \rightarrow 0$
\be
\label{bosonic2-limit} I_n^b  \rightarrow 2 (-1)^n T^{3-n}
\zeta(3-n)
\ee
 for $n \not= 2$. The $n-$even contributions to the effective 
action vanish anyway because the respective trace of
the $A$-fields equals zero, see Appendix \ref{traces}.


\subsection{Fermions}


The fermionic integrals of interest are
\be
 \label{fermionic1}
I^f_n(a, \Delta) = T^{3-n} \left( J^f_n(a,\Delta) + (-1)^{n+1}
J^f_n(-a,\Delta) \right)
 \ee
where $a \equiv \beta
\mu$ and
\be
J^f_n(a) = \int_\Delta^\infty dx \, x^2 \; \frac{d^n
f_{\rm FD}(x-a)}{dx^n} \;.
\ee
Expanding the Fermi-Dirac distribution
as
$$
f_{\rm FD} (x-a) = {1 \over e^{x-a} +1}=\sum_{m=1}^\infty
(-1)^{m-1} e^{-m (x-a)}  \ ,
$$
and interchanging the order of  summation and integration,
we obtain after performing the integral (\ref{integral})
\be \label{fermionic2}
J^f_n (a,\Delta) =  2 (-1)^{n-1}
\left( {\rm Li}_{3-n}(-e^{a-\Delta})
+ \Delta {\rm Li}_{2-n}(-e^{a-\Delta}) + \frac{\Delta^2}{2}
{\rm Li}_{1-n}(-e^{a-\Delta}) \right) \ .
\ee

Using the formula \cite{Actor:zf}
\be
{\rm Li}_N (-e^{-x}) = \sum_{n=0,}^\infty (-1)^{n+1}
\frac{x^n}{n!} \eta(N-n) \;.
\ee
where $\eta (s)$ is the alternating zeta function defined as
\be
\label{a-zeta} \eta (s)  \buildrel \rm def \over =
\sum_{m=1}^\infty {(-1)^{m-1} \over m^s} = (1 - 2^{1-s}) \:
\zeta(s) \;, \qquad {\rm Re}s > 1\;
\ee
we find that
\be \label{fermionic3}
J^f_n(a,\Delta) = 2 (-1)^n \sum_{l=0}^\infty
{a^l \over l!} \: \eta (3 -n -l) + {\cal O}(\Delta) \;,
\ee
when $\Delta \rightarrow 0$.

Inserting the series (\ref{fermionic3}) into Eq.~(\ref{fermionic1})
one gets
\be
\label{fermionic4}
I^f_n(a,0) =  2 T^{3-n} \sum_{l=0}^\infty {a^l \over l!}\:
\eta (3 -n -l) \Big[(-1)^n - (-1)^l \Big] \;.
\ee

As seen in Eq.~(\ref{fermionic4}), the argument of $\eta$ is
always an even number for non-vanishing terms. Since $\eta
(-2k)=0$ for $k = 1, 2, ...$ the series in Eq.~(\ref{fermionic4})
terminates. Then, one observes that $I^f_n(a,0) = 0$ for $n \ge 4$,
while
\ba
 I^f_0(a,0) &=& T^3 \left( 4a\, \eta(2) + {2\over 3}a^3 \eta(0) \right)
= T^3 \left({\pi^2 \over 3}\, a + {1 \over 3} \, a^3\right) \;, \nonumber \\
I^f_1(a,0) &=& T^2 \left(-4 \, \eta(2) - 2a^2 \, \eta(0) \right)
= - T^2 \left({\pi^2 \over 3} + a^2\right) \;, \nonumber \\
I^f_2(a,0) &=& 4 a T \, \eta(0)
= 2a T\;, \nonumber \\
I^f_3(a,0) &=& -4 \, \eta(0)
= - 2 \;, \nonumber \\
\ea where $\eta(2) = \pi^2 /12$ and $\eta(0) = 1/2$.


\section{Adjoint Representation Traces}
\label{traces}


To compute the effective action of \eq\nr{LMf} and \eq\nr{LMg}
one needs to evaluate traces in the fundamental and
adjoint representations. We present here some useful
formulas which relate the traces of the fundamental and
adjoint representations.

First, it is easy to prove that the total symmetric traces of an
odd number of adjoint generators vanish. In order to prove this,
note that
 \be {\rm Tr}[{\cal A}_0^m] = {\rm Tr}[({\cal
A}_0^m)^T] \;,
 \ee
 where the superscript $T$ denotes
transposition. The adjoint representation of $SU(N_c)$ is real,
and  the generators obey $T_a^T = - T_a$. Therefore,
\be {\rm
Tr}[{\cal A}_0^m] = (-1)^m {\rm Tr}[{\cal A}_0^m] \;.
\ee Thus,
the trace vanishes for odd $m$.

Two other useful formulas are:
\ba {\rm Tr}{\cal A}_0^2 &=& 2 N_c {\rm Tr} A_0^2
\ ,\\
{\rm Tr}{\cal A}_0^4 & = & 6 ({\rm Tr} A_0^2)^2
+ 2 N_c {\rm Tr} A_0^4 \;.
\ea
For $N_c=2$ and $N_c =3$, one also has
\be
{\rm Tr} A_0^4 = \frac{1}{2} ({\rm Tr} A_0^2)^2 \ .
\ee


\section{The one-loop bosonic action}
\label{one-loop}


In this appendix we show how the solutions of the transport
equations allow one to reproduce the results of
\cite{Gross:1980br,Weiss:1980rj,cka}. We consider only
the bosonic case for the $SU(2)$ gauge group. The fermionic
one can be treated similarly. Due to a global color rotation
a constant background field $A_0$ can be always chosen in
the diagonal form $A_0^a = \delta^{a3} C/g$ where $C$ is
a real constant. The respective gauge field in the adjoint
representation then reads
$$
{\cal A}^0 = \left( \begin{array}{ccc}
0 & - i C/g & 0 \\
i C/g & 0 & 0 \\
0 & 0 & 0
\end{array} \right) \ .
$$
Now, we plug this expression in \eq\nr{L-g} and expand the logarithm
and the exponential. Taking the trace ( ${\rm Tr} {\cal A}_0^{2m} =
2 (C/g)^{2m}$,  ${\rm Tr} {\cal A}_0^{2m+1} = 0$), we end up with
a series which after resumming reads
\be
\label{L-g-constant}
{\cal L}_g = {  T  \over \pi^2} \;  \int_0^{\infty} dE
\,E^2 \; \left[ {\rm ln}\big(1- e^{-\beta (E -C)} \big) +
{\rm ln}\big(1- e^{-\beta (E +C)} \big)
\right]  \;.
\ee
Rotating to  Euclidean time
${\cal A}_0 \rightarrow i  {\cal A}^E_0$, and thus
$C \rightarrow iC^E$, we find
\be
\label{LE-g-constant}
{\cal L}^E_g = {  T  \over \pi^2} \;  \int_0^{\infty} dE
\,E^2 \; \left[ {\rm ln}\big(1- e^{-\beta (E -iC^E)} \big) +
{\rm ln}\big(1- e^{-\beta (E +iC^E)} \big)
\right]
 \;,
\ee
 which matches the result found in Eq. (23) of
\cite{Weiss:1980rj}. The integral is evaluated in
\cite{Weiss:1980rj}, and non-analytic cubic terms in $C^E$ are
found. When rotated back to the real time these terms provide
imaginary contributions to the effective action. Its presence in
the Minkowski effective action could have been anticipated from
\eq\nr{L-g-constant}, as for  soft energies $E < C$ the
logarithm becomes a multi-valued complex function. The real time
integral is not well-defined and an additional prescription, such
as going to  Euclidean time, is necessary in order to evaluate it.
With the infrared cut-off $\Delta > C$, these problems are absent,
and the Minkowski effective action is real.


\end{document}